\documentclass[runningheads]{llncs}
\usepackage[T1]{fontenc}
\usepackage{graphicx}
\usepackage{xcolor}
\usepackage{booktabs}
\usepackage{amsmath,amsfonts}
\usepackage{hyperref}

\definecolor{notecolor}{RGB}{200,0,0}
\definecolor{outlinecolor}{RGB}{0,0,200}

\begin{document}
\title{BreastSegNet: Multi-label Segmentation of Breast MRI}
\titlerunning{BreastSegNet}

\author{Qihang Li\inst{1} \and
Jichen Yang\inst{1} \and
Yaqian Chen\inst{2} \and
Yuwen Chen \inst{2} \and
Hanxue Gu \inst{2} \and \\
Lars J. Grimm \inst{3} \and
Maciej A. Mazurowski \inst{1234}}

\authorrunning{Q. Li et al.}
\institute{Department of Biostatistics \& Bioinformatics, Duke University \and
Department of Electrical and Computer Engineering, Duke University \and
Department of Radiology, Duke University Medical Center \and
Department of Computer Science, Duke University}
% %
\maketitle              % typeset the header of the contribution
\begin{abstract}
Breast MRI provides high-resolution imaging critical for breast cancer screening and preoperative staging. However, existing segmentation methods for breast MRI remain limited in scope, often focusing on only a few anatomical structures, such as fibroglandular tissue or tumors, and do not cover the full range of tissues seen in scans. This narrows their utility for quantitative analysis. In this study, we present \textbf{BreastSegNet}, a multi-label segmentation algorithm for breast MRI that covers nine anatomical labels: fibroglandular tissue (FGT), vessel, muscle, bone, lesion, lymph node, heart, liver, and implant. We manually annotated a large set of 1123 MRI slices capturing these structures with detailed review and correction from an expert radiologist. Additionally, we benchmark nine segmentation models, including U-Net, SwinUNet, UNet++, SAM, MedSAM, and nnU-Net with multiple ResNet-based encoders. Among them, nnU-Net ResEncM achieves the highest average Dice scores of \textbf{0.694} across all labels. It performs especially well on heart, liver, muscle, FGT, and bone, with Dice scores exceeding \textbf{0.73}, and approaching \textbf{0.90} for heart and liver. All model code and weights are publicly available, and we plan to release the data at a later date.

\end{abstract}

\section{Introduction}
Breast magnetic resonance imaging (MRI) is one of the primary imaging modalities for breast cancer staging and high-risk screening, providing  superior contrast resolution and high sensitivity for early disease identification \cite{mann2008breast}. Multiple studies have recently demonstrated its superiority in breast cancer risk prediction, early detection, and diagnostic evaluation \cite{chen2025breast,wekking2023breast}, compared to traditional imaging modalities such as mammography. 

Despite these benefits, quantitative research on breast MRI remains limited, largely due to the lack of publicly available segmentation methods \cite{gubern2014breast}. A high-accuracy segmentation model that provides rich semantic information can facilitate a variety of quantitative research, including the analysis based on breast composition \cite{boyd2010breast} and registration \cite{chen2025guidedmorph}.

General-purpose medical image segmentation models, such as TotalSegmentator \cite{akinci2025totalsegmentator}, overlook labels specific to breast MRI, like fibroglandular tissue (FGT). A previously proposed breast MRI segmentation model proposed by Lew et al. \cite{lew2024publicly,buda2020data} provides only FGT, vessel, and breast labels. The lack of necessary labels precludes the direct extraction of important body composition parameters—such as muscle quality and bone density—from breast MRI \cite{mcgregor2014not,shepherd2017body}, thereby reducing the clinical applicability of these models.

To mitigate this gap, this study aims to provide a publicly available segmentation model and dataset that offers nine labels, including FGT, vessel, muscle, bone, lesion, lymph node, heart, liver, and implant. The main contributions to this work are:

\begin{enumerate}
\item We provide an automated, multi-class breast MRI segmentation model with nine anatomical labels, evaluated by the Dice similarity coefficient.
\item We developed and evaluated nine segmentation architectures, with the top-performing model, attaining an average Dice score greater than \textbf{0.694} for all nine labels.
\item We release the model code and pretrained weights on \url{https://github.com/mazurowski-lab/BreastSegNet}. We also describe the a breast MRI dataset and corresponding annotations which we plan to release at a later date.
\end{enumerate}

\section{Dataset}
\subsection{Data Selection}
\begin{figure}[!t]
    \centering
    \includegraphics[width=\columnwidth]
    {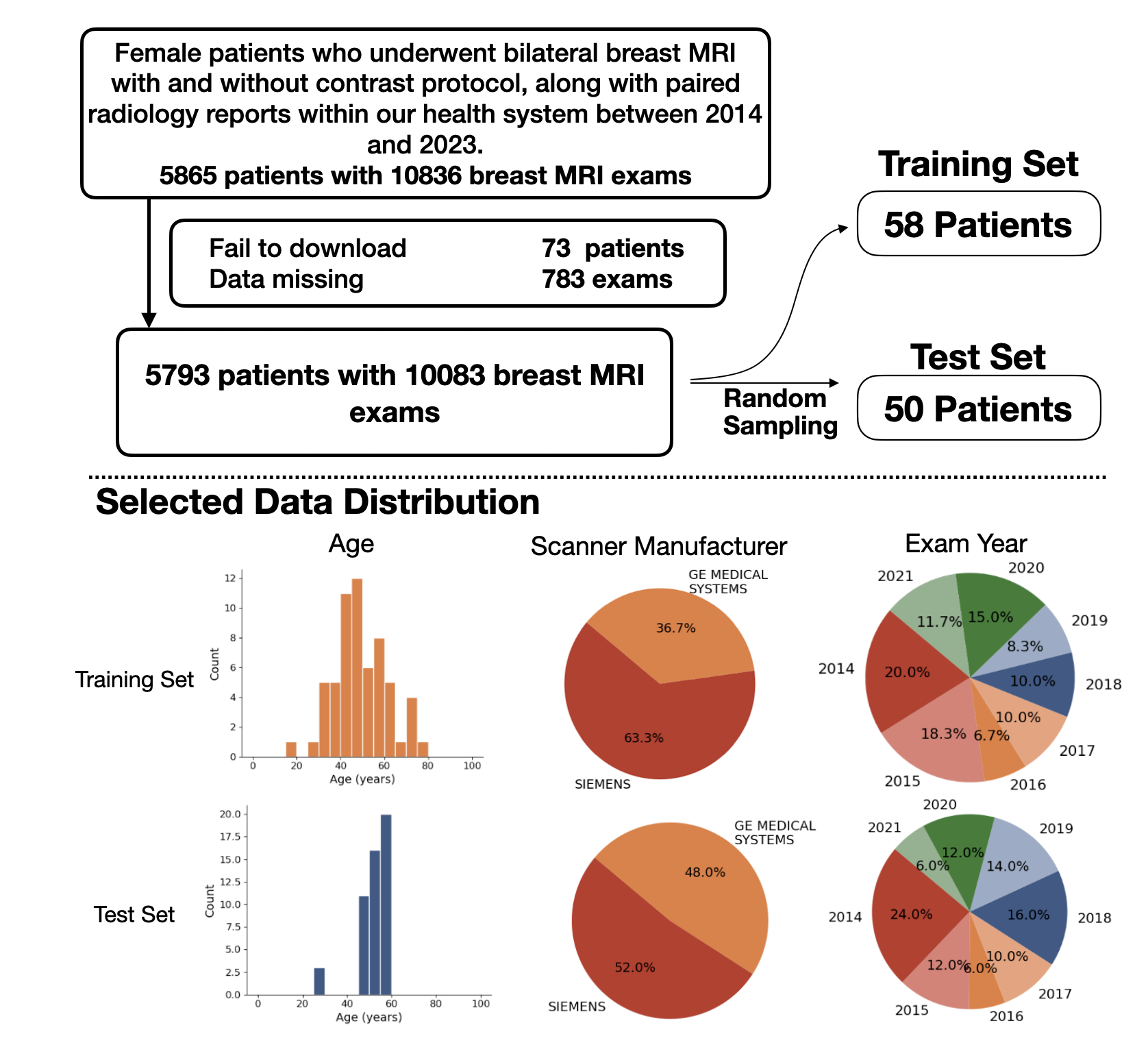}\\
    % {images/patient_population.jpg}\\
    \caption{The source data population, and the selected data distribution. }
    \label{fig:pat_pop}
\end{figure}

The initial cohort was collected from female patients who underwent bilateral breast MRI with and without contrast, along with paired radiology reports within our health system between 2014 and 2023. In total, we identified 5,865 patients with 10,836 breast MRI exams.
We then downloaded and de-identified these exams, and those that failed to download or were missing in the system were excluded. After this process, a total of 10,083 exams from 5,793 patients were available for further use.

From the initial cohort, we randomly selected 108 exams from different patients and divided them into two groups of 58 and 50 for training and testing, respectively. These volumes were identified manually. The data distribution for the training and testing sets is illustrated in Figure \ref{fig:pat_pop}.

\subsection{Annotation Pipeline}
\begin{figure}[!t]
    \centering
    \includegraphics[width=\columnwidth]{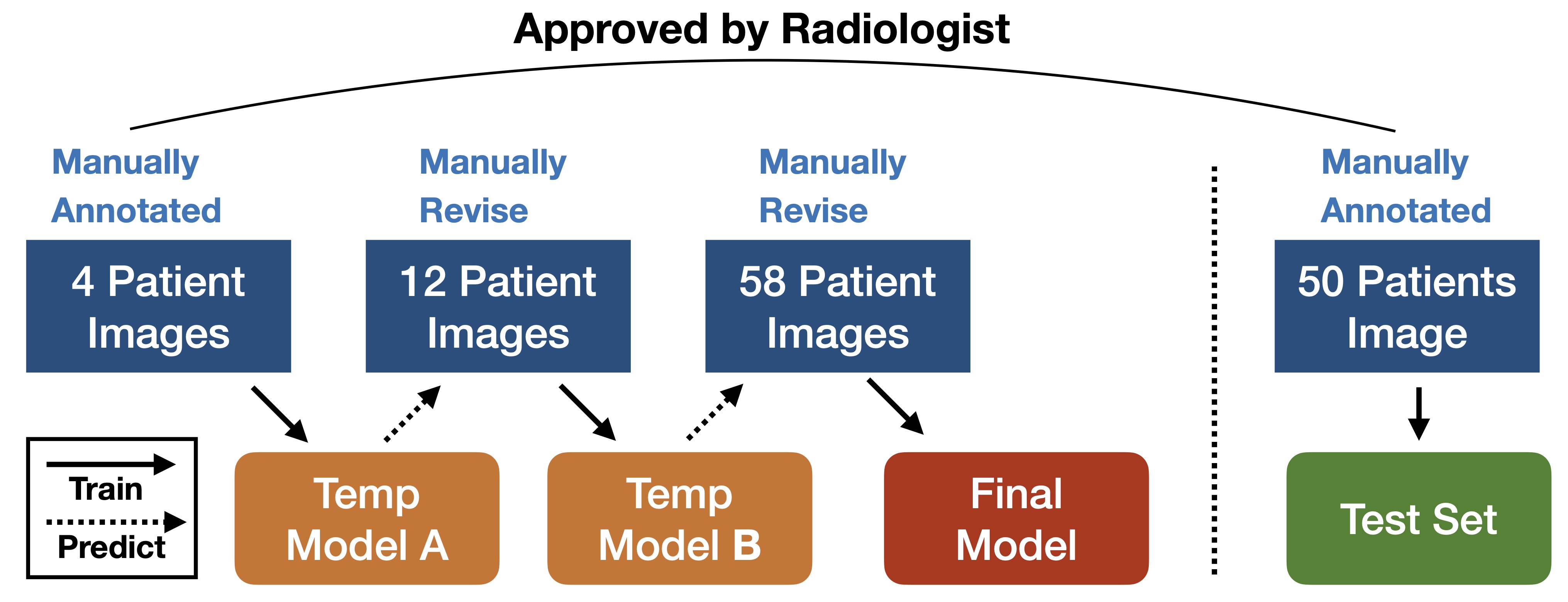}\\
    \caption{The data annotation pipeline. }
    \label{fig:annt_pl}
\end{figure}

The whole annotation process was carried out by four researchers without formal radiology training and one expert fellowship-trained breast radiologist who supervised the process. As illustrated in Figure \ref{fig:annt_pl}, we employed an iterative, model-assisted annotation workflow for breast MRI image annotation, with radiologist oversight at each stage. The annotation contains nine labels: tissue, vessel, muscle, bone, lesion, lymph node, heart, liver, and implant. To maximize the distinction between different labels, we selected the first post-contrast MRI sequence for annotation. The process consisted of multiple rounds of model development and manual refinement:

Initially, four patient MRIs were manually annotated by trained annotators, followed by radiologist review and approval to ensure accuracy. The annotations were applied every 10 slices. The annotators assigned each pixel to one of the nine labels, or background, to the best of their knowledge. The radiologist reviewed each annotated slice and provided feedback if they found any annotations inaccurate. Each slice was iteratively revised until the radiologist approved the annotation. These annotated cases were used to train a preliminary segmentation model (Temp Model A), which provided initial predictions for the next set of images.

Subsequently, predictions from Temp Model A were generated for 12 additional patient MRIs. These predictions were manually revised by annotators and reviewed by a radiologist to produce refined annotations. The updated dataset was then used to train a second model (Temp Model B).

In the final iteration, Temp Model B was applied to 58 new patient MRIs. The resulting segmentations were again manually corrected by researchers and approved by the radiologist. These curated annotations were used to train the final model.

For independent evaluation, a separate test set of 50 patient MRIs was manually annotated without model assistance. For each volume, 3 random slices were randomly selected from the breast region and annotated. All annotations in this test set underwent radiologist review to serve as the ground truth for model performance assessment.

\subsection{Segmentation Details}
% Unlike previous studies with limited labels for breast MRI, this work annotates nine distinct types of body composition to classify every possible pixel in breast MRI scans. This section outlines the annotation process for the following labels: FGT, vessel, muscle, bone, lesion, lymph node, heart, liver, and implant.

This section provides detailed definitions of the nine labels supported by our model, including FGT, vessel, muscle, bone, lesion, lymph node, heart, liver, and implant.

\begin{figure}[!t]
    \centering
    \includegraphics[width=\columnwidth]{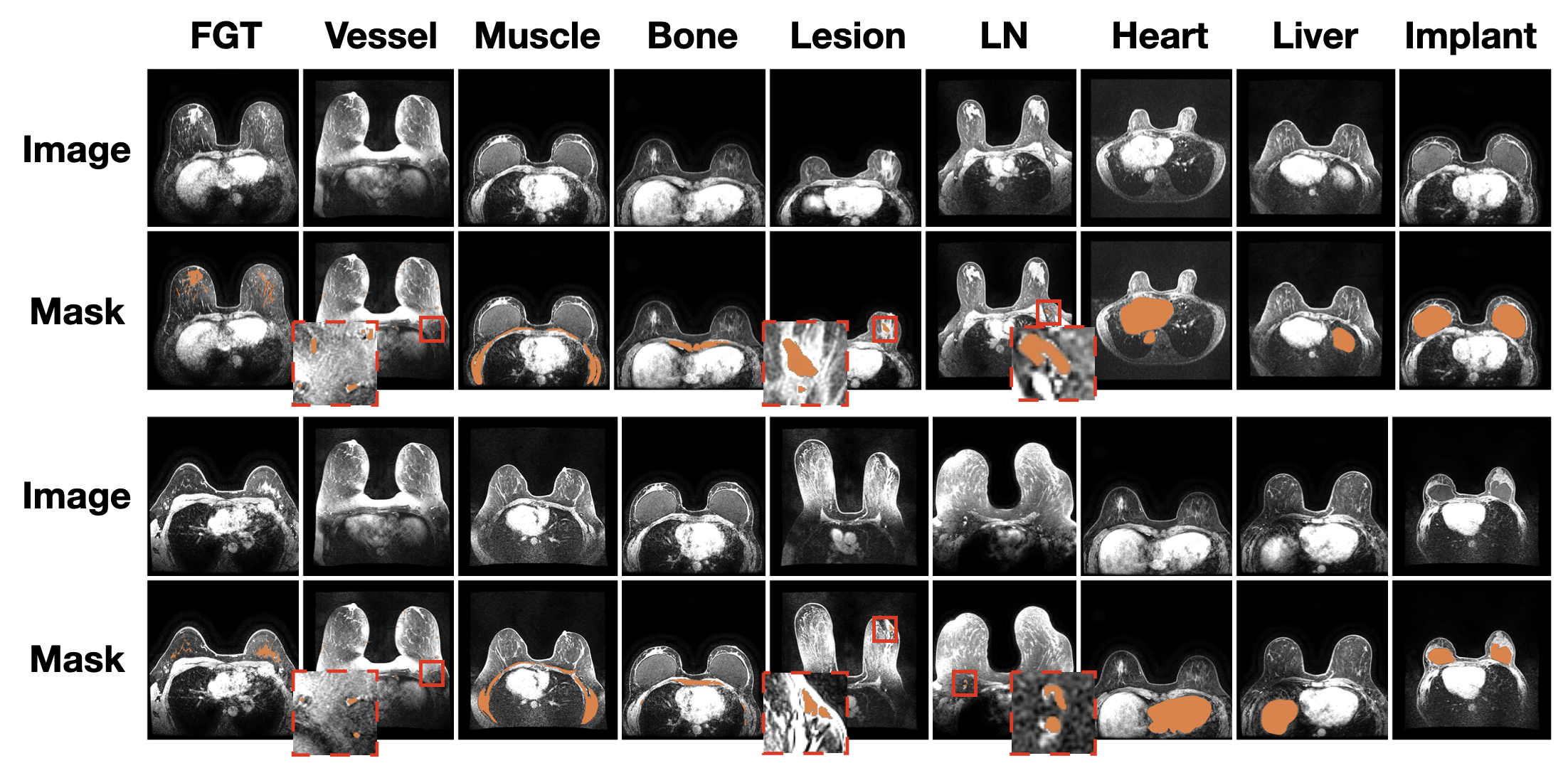}\\
    \caption{\textbf{Annotation visualization for nine anatomical labels:} For each label, two representative slices are shown: the original image (rows 1 and 3) and its corresponding segmentation overlaid in \textit{orange} (rows 2 and 4). For small or sparse structures (Vessel, Lesion, LN), zoomed-in views are provided in dashed red boxes, with the original regions marked by solid red boxes.}
    \label{fig:seg_sample}
\end{figure}

The FGT, vessel, muscle, bone, and lymph node, annotations strictly adhered to anatomical definitions, as these classifications present no inherent ambiguity. Lesion refers to all abnormal findings within the breast that are not artificially implanted, encompassing masses, non-mass enhancements, and metastatic/ cancerous lymph nodes. Due to the presence of artifacts, the boundaries of the heart and liver are often indistinct on MRI, but we have made every effort to differentiate them. In this annotation, implant specifically denotes materials used in breast augmentation procedures, including saline or silicone implants. Other artificial objects, such as implantable infusion ports, were not labeled.

Figure 3 presents example slices with overlaid annotations (\textit{orange}) for the nine segmented labels. For each label, two randomly selected slices are shown. The first two rows correspond to the original image and the image with overlaid segmentation mask from the first selected slice, while the third and fourth rows show the same for a second slice. 

\section{Methods}
\subsection{Segmentation methods}
To select the best-performing algorithm for our task, we explored nine segmentation methods, which include U-Net, SwinUNet, U-Net++, fine-tuned SAM, fine-tuned MedSAM, nnU-Net, nnU-Net ResEncM, nnU-Net ResEncL, and nnU-Net ResEncXL. These methods are grouped into three categories for better comparison: (1) the U-Net-based category, which includes U-Net, SwinUNet, and U-Net++; (2) the foundation model category, comprising fine-tuned SAM and fine-tuned MedSAM; and (3) the nnU-Net-based category, consisting of nnU-Net and its variants with ResEncM, ResEncL, and ResEncXL encoder \cite{cao2022swin,nnunet,revistnnunet,SAM,medsam,unet,zhou2018unet++}. For detailed descriptions of the models in each category, please refer to Appendix \ref{model_detail}.

\subsection{Evaluation metrics}
We employed the Dice coefficient as the metric to evaluate the model. Dice is one of the most commonly utilized metrics, measuring the image overlap between predicted and ground truth regions, and reflecting the spatial agreement of segmentation boundaries. 

The formula for the Dice coefficient is Eq. \eqref{Eq:Dice}, where A represents the model-predicted mask and the B represents the set of pixels in the ground truth. Dice ranges from 0 to 1, where a value of 1 indicates perfect overlap and 0 signifies no overlap \cite{dice1945measures}.
\begin{equation}\label{Eq:Dice}
    \text{Dice} = \frac{2 |A \cap B|}{|A| + |B|}
\end{equation}

\section{Experiments and Results}
\subsection{Implementation Details}
This study utilized 58 volumetric datasets from 58 patients for training. Of these, 812 slices from 48 volumes were allocated to the training set and 161 slices from 10 volumes to the validation set.

All models were configured for 2D prediction. For the nnU-Net series, default slice-wise normalization was applied. To ensure consistency, slice-wise normalization was also used for the other models. The fine-tuned models, including SAM and MedSAM, were trained for 100 epochs on our datasets. All other models were trained for a total of 1000 epochs, and the model achieving the best performance on the validation set was selected as the final model. All models converged during training.

\subsection{Results}
\begin{figure}[!t]
    \centering
    \includegraphics[width=\columnwidth]{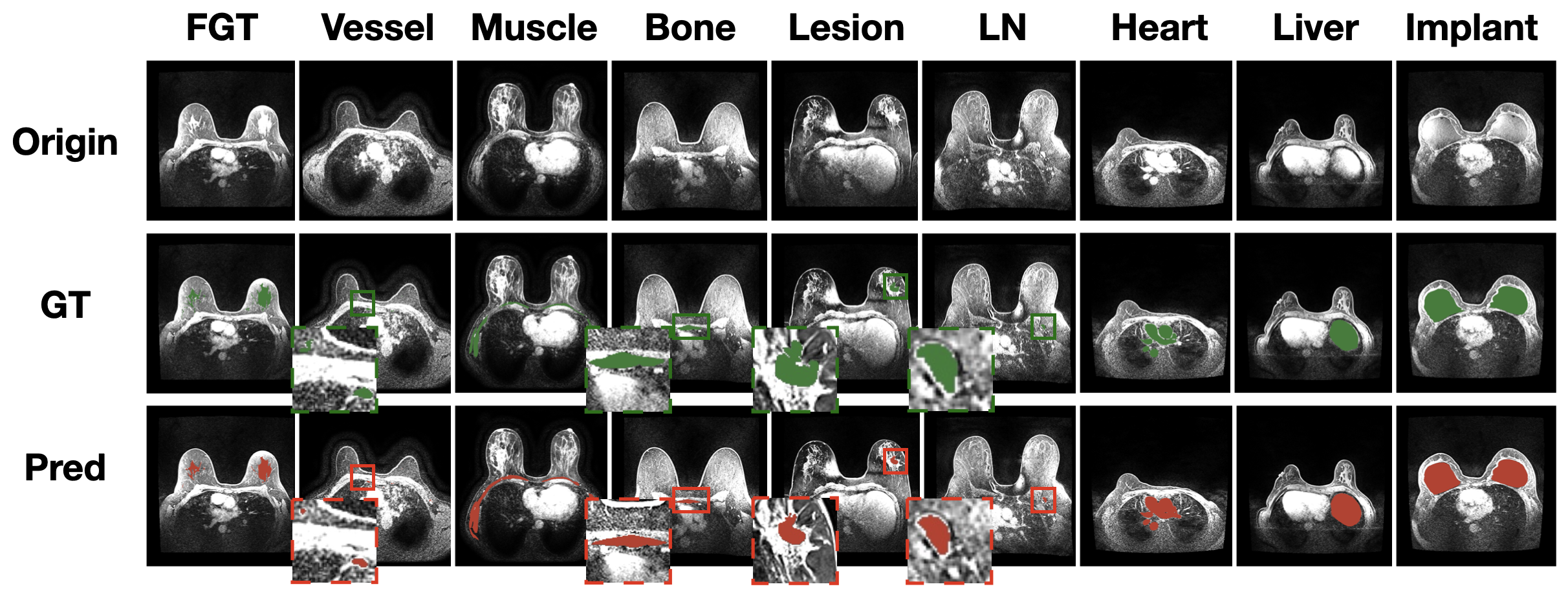}\\
    \caption{\textbf{Qualitative comparison:} Segmentation results of nnU-Net ResEncM, the best-performing model, compared to ground truth (GT) across nine anatomical labels. Rows show the original image, GT (\textit{green}), and prediction (\textit{red}). Zoomed-in views are shown for small structures (Vessel, Lesion, LN).}
    \label{fig:seg_results}
\end{figure}

To evaluate segmentation performance, we computed the average 2D Dice score across all slices containing non-zero labels for each anatomical label across the nine selected models. The quantitative results are detailed in Table \ref{tab:seg_results_baseline}, and a qualitative comparison between the prediction of nnU-Net ResEncM, which demonstrates the highest average performance, and the ground truth is shown in Figure \ref{fig:seg_results}.

As demonstrated in Table \ref{tab:seg_results_baseline}, nnU-Net and its ResEnc variants significantly outperform the models in the UNet and Foundation Model categories for this task, with the best-performing models belonging to nnU-Net and its ResEnc variants across all nine labels. Among the four models in the nnU-Net-based category, nnU-Net ResEncM exhibits the highest average performance, achieving an average Dice score of \textbf{0.694}.

Segmentation performance varies notably across anatomical labels, with some structures consistently achieving higher Dice scores. For example, all models in the nnU-Net-based category achieve Dice scores above \textbf{0.73} for the FGT and bone labels, over \textbf{0.8} for the muscle label, and approach \textbf{0.9} for the heart and liver labels. For some more challenging regions such as vessel, lesion, and implant, the performance varies significantly across the models, with the best scores reaching \textbf{0.58}, \textbf{0.66}, and \textbf{0.81}, respectively.

Notably, lymph nodes exhibit the lowest segmentation performance among the nine labels. We discuss potential reasons for this limitation and plans to address it in the discussion section.

For qualitative evaluation, Figure \ref{fig:seg_results} presents the comparison between ground truth (GT) and the predictions (Pred) from nnU-Net ResEncM across all nine anatomical labels. The original MRI slice is shown in the first row, followed by the GT mask (\textit{green}) and the predicted mask (\textit{red}). For large and clearly defined structures such as FGT, muscle, bone, heart, and liver, the model shows strong agreement with the ground truth, with minimal boundary discrepancy. For smaller and more challenging labels like vessel, lesion, and lymph node, visual differences become more apparent. For these small or sparse regions, zoomed-in boxes are provided to better assess model performance.

\begin{table}[h]
\centering
\caption{Segmentation results for nine models, evaluated by Dice ($\uparrow$). The highest score for each label among all models is highlighted in bold.The right‑most column reports the mean Dice for each label across every model. Acronyms: FGT – fibroglandular tissue.}
\label{tab:seg_results_baseline}
\vspace{5pt} % Adds a little space between caption and table
% Adjusted for 5 models * 1 metric = 5 'c' columns + 1 'l' column
\begin{tabular*}{\linewidth}{@{\extracolsep{\fill}}l ccccc @{}}
\toprule % Top thick horizontal line
\multicolumn{6}{c}{\textbf{Models in UNet and Foundation Model Categories}} \\
\midrule
\textbf{Composition} & \textbf{UNet} & \textbf{SwinUNet} & \textbf{UNet++} & \textbf{SAM} & \textbf{MedSAM} \\
\midrule % Line after header
\textbf{FGT} & 0.66 & 0.69 & 0.69 & 0.60 & 0.56\\
\textbf{Vessel} & 0.41 & 0.46 & 0.50 & 0.23 & 0.18\\
\textbf{Muscle} & 0.68 & 0.69 & 0.74 & 0.70 & 0.64\\
\textbf{Bone} & 0.61 & 0.62 & 0.67 & 0.51 & 0.42\\ 
\textbf{Lesion} & 0.00 & 0.01 & 0.07 & 0.05 & 0.01\\
\textbf{Lymph node} & 0.00 & 0.00 & 0.03 & 0.14 & 0.05\\
\textbf{Heart} & 0.85 & 0.85 & 0.83 & 0.86 & 0.84\\
\textbf{Liver} & 0.42 & 0.25 & 0.26 & 0.71 & 0.54\\
\textbf{Implant} & 0.60 & 0.18 & 0.53 & 0.80 & 0.74\\
\midrule % Separates the individual results from the overall summary
\textbf{Overall} & 0.470 & 0.417 & 0.480 & 0.511 & 0.442\\
\toprule % Top thick horizontal line
\multicolumn{6}{c}{\textbf{Models in nnU-Net-based Category}}\\
\midrule
\textbf{Composition} & \textbf{nnU-Net} & \textbf{ResEncM} & \textbf{ResEncL} &  \textbf{ResEncXL} & \textbf{Avg}\\
\midrule % Line after header
\textbf{FGT} & 0.74 & \textbf{0.75} & 0.74 & 0.74 & 0.69\\
\textbf{Vessel} & \textbf{0.58} & 0.57 & 0.56 & 0.57 & 0.45\\
\textbf{Muscle} & 0.80 & \textbf{0.81} & 0.80 & 0.81 & 0.74\\
\textbf{Bone} & \textbf{0.73} & \textbf{0.73} & \textbf{0.73} & \textbf{0.73} & 0.64\\
\textbf{Lesion} & 0.38 & \textbf{0.66} & 0.38 & 0.63 & 0.24\\
\textbf{LN} & 0.13 & 0.12 & \textbf{0.16} & 0.10 & 0.08\\
\textbf{Heart} & \textbf{0.92} & 0.91 & 0.91 & 0.91 & 0.88\\
\textbf{Liver} & 0.89 & \textbf{0.90} & \textbf{0.90} & \textbf{0.90} & 0.64\\
\textbf{Implant} & \textbf{0.81} & 0.80 & 0.76 & 0.69 & 0.66\\
\midrule % Separates the individual results from the overall summary
\textbf{Overall} & 0.664 & \textbf{0.694} & 0.660 & 0.675 & 0.56\\
\bottomrule % Bottom thick horizontal line
\end{tabular*}
\end{table}

\section{Discussion}
In this study, we proposed \textbf{BreastSegNet}, a multi-label segmentation network for breast MRIs, to address the limitations of prior models, which typically target limited structures. By incorporating nine anatomical labels, including clinically important regions such as fibroglandular tissue, vessel, muscle, bone, lesion, lymph node, heart, liver, and implant, our model enables more comprehensive downstream analysis in breast imaging. To ensure that our data results in the most accurate model, evaluate nine segmentation frameworks, with nnU-Net ResEncM demonstrating the highest Dice score of 0.694 averaged across all nine anatomical labels. Notably, segmentation accuracy is especially strong for large anatomical structures such as the heart, liver, muscle, FGT, and bone, each achieving Dice scores above 0.73 and approaching 0.90 for the heart and liver.

The limitation of our model is the lower segmentation performance for smaller or less visually distinct structures, especially lymph nodes. We identify three key factors contributing to this lower performance: (1) Lymph nodes are rare. Only 15 out of 150 test slices have lymph nodes. This strong data imbalance makes the model prone not to predict them, as the original loss treats matched absence in prediction and ground truth as zero loss. (2) Lymph nodes have similar attentuation values as blood vessels due to their vascular supply, making it easy for the model to confuse them. (3) Lymph nodes, especially non-cancerous lymph nodes, are usually small structures, so slight mislabeling in prediction leads to a dramatic Dice drop.

\clearpage
\bibliographystyle{splncs04}
\bibliography{refs}

\appendix
\section{Model Architecture Details}\label{model_detail}
\subsubsection{U-Net} U-Net is one of the most commonly used benchmarks for medical image segmentation tasks. Compared to traditional convolutional networks, U-Net features a symmetric encoder–decoder structure that helps preserve spatial resolution during upsampling. SwinUNet builds on the original U-Net architecture by integrating the Swin Transformer into the encoder, enabling the capture of long-range dependencies. U-Net++, on the other hand, introduces nested and dense skip connections, which help reduce the semantic gap between encoder and decoder features.
 
\subsubsection{Foundation model} Foundation model fine-tuning has become a recent trend in medical segmentation tasks due to its strong adaptability and generalizability. It has demonstrated state-of-the-art performance across multiple medical segmentation tasks \cite{gu2025segmentanybone}. Among these models, the Segment Anything Model (SAM) and MedSAM are two of the most commonly used foundation models for medical image segmentation. SAM is trained on a large corpus of natural images and demonstrates strong zero-shot segmentation capabilities. MedSAM is further adapted for medical applications by being pretrained on a diverse set of medical imaging data, making it widely applicable across segmentation tasks.

\subsubsection{nnU-Net-based} We separate the nnU-Net algorithms from the U-Net category to highlight the performance differences introduced by different nnU-Net planners. The original nnU-Net extends the U-Net architecture by incorporating automated configuration of preprocessing, architecture adaptation, training, and postprocessing pipelines, demonstrating significant performance improvements over the vanilla U-Net. In its recent update, nnU-Net further introduces enhancements to the U-Net baseline, emphasizing the importance of using advanced CNN architectures such as ResNet. The ResEncM, ResEncL, and ResEncXL planners represent nnU-Net variants with progressively deeper ResNet encoder backbones—ResNet-34, ResNet-50, and ResNet-101.
\end{document}